\documentclass[aps,pra,twocolumn,superscriptaddress,nofootinbib]{revtex4-1}

\usepackage{amsmath}
\usepackage{amssymb}
\usepackage{graphicx}
\usepackage{mathrsfs}
\usepackage{ntheorem}
\usepackage{mathtools}
\usepackage{enumerate}
\usepackage{enumitem}
\usepackage{times,txfonts}
\usepackage{subfigure}
\usepackage{upgreek}
\usepackage{tikz}
\usepackage[english]{babel}

\usepackage[
    colorlinks,
    bookmarksopen,
    bookmarksnumbered,
    citecolor=blue,
    linkcolor=blue,
    urlcolor=blue
]{hyperref}

\newcommand{\ket}[1]{|#1\rangle}

\def\CC{{\rm\kern.24em
  \vrule width.04em height1.46ex depth-.07ex
  \kern-.30em C}}

\def\RR{{\rm\kern.24em
  \vrule width.04em height1.46ex depth-.07ex
  \kern-.30em R}}

\def\P{{\rm I\kern-.25em P}}

\begin{document}

\lefthyphenmin=2
\righthyphenmin=3

\title{No universal LOCC entanglement purification}

\author{C. L. Liu}
\email{clliu@sdu.edu.cn}
\affiliation{School of Information Science and Engineering, Shandong University, Qingdao 266237, China}
\author{Baoqing Sun}
\email{baoqing.sun@sdu.edu.cn}
\affiliation{School of Information Science and Engineering, Shandong University, Qingdao 266237, China}
\author{D. M. Tong}
\email{tdm@sdu.edu.cn}
\affiliation{School of Physics, Shandong University, Jinan 250100, China}
\date{\today}

\begin{abstract}
Entanglement purification aims to obtain high-fidelity entangled states
from noisy entangled states. This task becomes particularly challenging
when the available input states are finite in number, only partially
characterized, and may not be identical.
Whether there exists a universal purification protocol applicable to such inputs has therefore been a longstanding question. This motivates the study of universal $n$-to-$1$ entanglement purification. Although no such LOCC protocol exists for the full set of two-qubit entangled states, it has remained open whether this no-go result extends to higher-dimensional bipartite states and multipartite states.  In this work, we resolve
this question by proving that, for any fixed finite local dimensions, no $n$-to-$1$ LOCC protocol is universal for all bipartite entangled
states or all multipartite entangled states. We further show that the
bipartite and multipartite no-go results persist under several natural
extensions and establish an analogous no-go theorem for coherence
purification under strictly incoherent operations.
\end{abstract}

\maketitle

\section{Introduction.}
Entanglement is a fundamental resource for a wide range of quantum
information processing tasks, including quantum communication,
distributed quantum computation, and quantum cryptography
~\cite{Nielsen2000,Briegel1998,Wehner2018,Azuma2023,Horodecki2009,
Aolita2015}. However, noise arising during entanglement generation,
distribution, and storage typically degrades shared entangled states.
Entanglement purification protocols (EPPs)~\cite{note} were therefore
introduced to distill high-fidelity entangled states from noisy inputs
using local operations and classical communication (LOCC)
~\cite{Bennett1996b,Bennett1996,Deutsch1996}.
Many EPPs, however, are tailored to specified input-state families and
assume access to independently and identically distributed
(i.i.d.) copies. In realistic scenarios, only finitely many
incompletely characterized input states may be available, and they
may not be identical.
These limitations motivate two
longstanding objectives in entanglement purification: reducing the
prior information required about the inputs and dispensing with the
assumption that i.i.d.\ copies are available
~\cite{Dur2007,Yan2023}.

Progress toward the first objective includes the Schmidt projection
method of entanglement concentration, which requires no prior
knowledge of the Schmidt coefficients but assumes identically
prepared pure-state inputs~\cite{Bennett1996Concentration}, and EPPs
for unknown or partially known states under an exchangeability
assumption~\cite{Brun2001}. Universal protocols~\cite{note2} include
distortion-free entanglement concentration for unknown bipartite pure
states~\cite{Matsumoto2007} and restricted classes of multipartite
pure states~\cite{Botero2018}, as well as asymptotic entanglement
distillation from unknown i.i.d.\ copies under non-entangling
operations~\cite{Lami2026}.
Progress toward relaxing the assumption of i.i.d.\ copies
includes optimal single-copy probabilistic transformations of
bipartite pure states~\cite{Nielsen1999,JonathanPlenio1999,Vidal1999} and entanglement enhancement
and local filtering for two-qubit states
~\cite{Kent1999,Verstraete2001,Kwiat2001}. General one-shot formulations have
further characterized nonasymptotic entanglement distillation under
LOCC and broader classes of operations
~\cite{Fang2019,Regula2019,Ecker2021}. The universal results above retain the i.i.d.\ assumption, whereas the
finite-resource results apply only to particular input classes or
under additional assumptions. Thus, the two objectives have largely
been pursued separately. This raises a natural question: does there exist a universal LOCC EPP
applicable to finitely many, possibly nonidentical input states when
only limited prior information is available?

More recently, this question has been investigated within the
framework of universal $n$-to-$1$ LOCC EPPs
~\cite{Zang2025,Kalman2025,Zhou2025,Orthey2026,Zhao2026}. Given $n$
input states $\rho_1,\ldots,\rho_n$, let $\Lambda_0$ denote the
successful branch of an $n$-to-$1$ LOCC EPP acting on their total
state $\rho^{\mathrm{in}}=\bigotimes_{j=1}^{n}\rho_j$.
Such an EPP is universal if, for every
admissible choice of the input states, the success probability
$p_{\rho^{\mathrm{in}}}
=\operatorname{Tr}[\Lambda_0(\rho^{\mathrm{in}})]$ is positive and the
normalized successful output
$\rho'=\Lambda_0(\rho^{\mathrm{in}})/p_{\rho^{\mathrm{in}}}$ satisfies
$F(\rho')\geq\max_{1\leq j\leq n}F(\rho_j)$, where all fidelities are
evaluated with respect to the same prescribed target state~\cite{Zang2025}. The same work also
proved that no such EPP exists for the full set of two-qubit entangled
states. However, it has remained open whether this no-go result extends, at fixed finite local dimensions, to all higher-dimensional bipartite entangled states and all multipartite entangled states.

In this work, we resolve the above question. Specifically, we show that for every $n\geq2$ and any fixed finite
local dimensions, no $n$-to-$1$ LOCC EPP is universal for
either the full set of bipartite entangled states or the full set of
multipartite entangled states.  Moreover, we show that the bipartite and
multipartite no-go results remain valid under separable operations,
when the allowed inputs are restricted to states with fidelities
arbitrarily close to unity, and with the assistance from a
finite-dimensional pure catalyst. Beyond entanglement, we establish an
analogous no-go theorem for coherence purification under strictly
incoherent operations.

\section{Preliminaries.}
Let us consider $n\geq2$ entangled states
$\rho_1$, $\ldots$, $\rho_n$ shared by $m\geq2$ spatially separated parties,
denoted by $\mathcal P=\{A,B,C,\ldots\}$. Let
$\mathcal H=\bigotimes_{X\in\mathcal P}\mathcal H_X$, where
$\mathcal H_X$ is the finite-dimensional local Hilbert space of party
$X$. Each $\rho_j$ acts on an isomorphic copy of $\mathcal H$, whose
local subsystem corresponding to party $X$ is denoted by $X_j$, with
$\mathcal H_{X_j}\cong\mathcal H_X$. Their tensor product
$\rho^{\mathrm{in}}:=\bigotimes_{j=1}^{n}\rho_j$ is the total input
state. For the output system, the local subsystem of party $X$ is
denoted by $X'$, with $\mathcal H_{X'}\cong\mathcal H_X$.
Let $\mathcal B(\mathcal K)$ denote the space of linear operators on a
Hilbert space $\mathcal K$. An $n$-to-$1$ EPP is a two-outcome LOCC quantum instrument
$\Lambda=\{\Lambda_0,\Lambda_1\}$~\cite{Bennett1996,Chitambar2014},
where
$\Lambda_0,~\Lambda_1:
\mathcal B(\mathcal H^{\otimes n})
$$\to$$\mathcal B(\mathcal H)$
are completely positive and trace-nonincreasing maps corresponding to
the success and failure branches, respectively, and
$\Lambda_0+\Lambda_1$ is trace preserving. For the total input state
$\rho^{\mathrm{in}}$, the success probability is
$p_{\rm s}=\operatorname{Tr}[\Lambda_0(\rho^{\mathrm{in}})]$.
Whenever $p_{\rm s}>0$, the normalized output conditioned on success is
$\rho'=\Lambda_0(\rho^{\mathrm{in}})/p_{\rm s}$.

We first define universality for EPPs acting on fixed finite-dimensional bipartite states. For $m=2$, let
$\mathcal H=\mathbb C^{d_1}\otimes\mathbb C^{d_2}$, with the two
factors associated with parties $A$ and $B$, respectively. For the
$j$th input state $\rho_j$, the corresponding local subsystems are
denoted by $A_j$ and $B_j$. Let $\mathcal S_{d_1,d_2}$ denote the set
of all entangled states on $\mathcal H$. Set $d=\min\{d_1,d_2\}$ and choose a maximally
entangled target state of Schmidt rank $d$, denoted by
$|\Phi_d\rangle$. Maximally entangled states are natural targets for
purification because they are key resources for various quantum information processing tasks, such as quantum teleportation,
quantum networks, and quantum communication
~\cite{Nielsen2000,Horodecki2009}.
In suitable local bases, the target state can be written as
$|\Phi_d\rangle=d^{-1/2}\sum_{r=0}^{d-1}|r\rangle_A|r\rangle_B$,
and we denote
$\Phi_d:=|\Phi_d\rangle\langle\Phi_d|$. For any state $\sigma$ on
$\mathcal H$, the fidelity~\cite{Jozsa1994} with the target state is
$F(\sigma):=\langle\Phi_d|\sigma|\Phi_d\rangle$.
Generalizing the definition introduced for $2\otimes2$ entangled
states in Ref.~\cite{Zang2025}, we call an $n$-to-$1$ EPP
$\{\Lambda_0,\Lambda_1\}$ universal for
$\mathcal S_{d_1,d_2}$ if, for every total input state
$\rho^{\mathrm{in}}=\bigotimes_{j=1}^{n}\rho_j$ with
$\rho_j\in\mathcal S_{d_1,d_2}$ for all $j=1,\ldots,n$, the protocol
succeeds with probability
$p_{\rm s}(\rho^{\mathrm{in}})
:=\operatorname{Tr}[\Lambda_0(\rho^{\mathrm{in}})]>0$
and, conditioned on success, produces the normalized output state
$\rho'=\Lambda_0(\rho^{\mathrm{in}})/
p_{\rm s}(\rho^{\mathrm{in}})$ satisfying
$F(\rho')\geq\max_{1\leq j\leq n}F(\rho_j)$.

We next define universality for EPPs acting on genuinely multipartite
entangled states. For $m\geq3$, let $d_X\geq2$ denote the local dimension of
party $X\in\mathcal P$, and write
$\boldsymbol d=(d_A,d_B,d_C,\ldots)$. We take
$\mathcal H_X=\mathbb C^{d_X}$ for every $X\in\mathcal P$ and let
$\mathcal G_{\boldsymbol d}$ denote the set of genuinely multipartite
entangled states on
$\mathcal H=\bigotimes_{X\in\mathcal P}\mathbb C^{d_X}$.
Set $d=\min_{X\in\mathcal P}d_X$ and
$D=\prod_{X\in\mathcal P}d_X$. For each party $X\in\mathcal P$, let
$\{|r\rangle_X\}_{r=0}^{d_X-1}$ denote a suitable local basis.
As the target state, we choose the generalized GHZ state~\cite{Huber2010}
$|\Psi_{\boldsymbol d}\rangle=d^{-1/2}\sum_{r=0}^{d-1}
\bigotimes_{X\in\mathcal P}|r\rangle_X$, with
$\Psi_{\boldsymbol d}:=|\Psi_{\boldsymbol d}\rangle
\langle\Psi_{\boldsymbol d}|$. For any state $\sigma$
on $\mathcal H$, its fidelity with the target GHZ state is
$F(\sigma):=\langle\Psi_{\boldsymbol d}|\sigma|
\Psi_{\boldsymbol d}\rangle$. An $n$-to-$1$ EPP is universal for
$\mathcal G_{\boldsymbol d}$ if, for every total input state
$\rho^{\mathrm{in}}=\bigotimes_{j=1}^{n}\rho_j$ with
$\rho_j\in\mathcal G_{\boldsymbol d}$, the protocol succeeds with
positive probability and, conditioned on success, produces an output
state $\rho'$ satisfying
$F(\rho')\geq\max_{1\leq j\leq n}F(\rho_j)$.

\section{No universal LOCC entanglement purification.}
With the above
definitions in place, we first consider the bipartite case.

\textbf{Theorem 1.} For any finite local dimensions $d_1,d_2\geq2$
and any $n\geq2$, no $n$-to-$1$ LOCC EPP is universal for $\mathcal S_{d_1,d_2}$.

\textit{Proof.} We begin by considering the case $d_1=d_2=d$ and write
$\mathcal S_d:=\mathcal S_{d,d}$. Suppose, for contradiction, that a
universal $n$-to-$1$ LOCC EPP for $\mathcal S_d$ exists.

First, we construct input states for which universality requires the
normalized output conditioned on success to be maximally entangled.

For each pair $A_jB_j$, let
$|\Phi_d\rangle_j=d^{-1/2}\sum_{r=0}^{d-1}
|r\rangle_{A_j}|r\rangle_{B_j}$ and
$\Phi_d^{(j)}:=|\Phi_d\rangle_j\langle\Phi_d|$.
Define
$\tau_j=p\Phi_d^{(j)}+(1-p)\frac{I_{A_jB_j}}{d^2}$,
where $I_{A_jB_j}$ is the identity on
$\mathcal H_{A_j}\otimes\mathcal H_{B_j}$ and
$\frac{1}{d+1}<p<1$. For this range of $p$, every $\tau_j$ is full
rank and entangled, and hence $\tau_j\in\mathcal S_d$. We consider the
following $n+1$ total input states
\begin{eqnarray}
\Omega_0
=
\bigotimes_{j=1}^n\Phi_d^{(j)},
~~
\Omega_k
=
\Phi_d^{(k)}
\otimes
\bigotimes_{j\neq k}\tau_j,
~~
k=1,\ldots,n .
\label{eq:test-inputs}
\end{eqnarray}
Let $|\Phi_d\rangle_{\rm out}
=d^{-1/2}\sum_{r=0}^{d-1}|r\rangle_{A'}|r\rangle_{B'}$ denote the corresponding maximally
entangled target state on the output pair $A'B'$.
Since each $\Omega_k$ contains the target state $\Phi_d$ among its
$n$ input states, the maximum input fidelity is $1$.
Universality therefore requires the normalized output conditioned on
success to equal $\Phi_d^{\rm out}$. Equivalently,
\begin{eqnarray}
\Lambda_0(\Omega_k)
=
p_k\Phi_d^{\rm out},
\qquad
k=0,\ldots,n,
\label{eq:pure-output}
\end{eqnarray}
where
$p_k:=\operatorname{Tr}[\Lambda_0(\Omega_k)]>0$.
Since $\Lambda_0$ is the successful branch of an LOCC protocol, it
admits a product Kraus representation
$\Lambda_0(\rho)=\sum_\mu K_\mu\rho K_\mu^\dagger$, with
$K_\mu=M_\mu\otimes N_\mu$. Here,
$M_\mu:\bigotimes_{j=1}^n\mathcal H_{A_j}\to\mathcal H_{A'}$ and
$N_\mu:\bigotimes_{j=1}^n\mathcal H_{B_j}\to\mathcal H_{B'}$.

Second, we derive support constraints for every product Kraus operator
in the successful branch and then select one satisfying a nonzero
condition.

Let $\mathcal L_{\rm out}$ be the subspace spanned by
$|\Phi_d\rangle_{\rm out}$. Then
$P_\perp=I_{A'B'}-\Phi_d^{\rm out}$ is the orthogonal projector onto
$\mathcal L_{\rm out}^{\perp}$, where $I_{A'B'}$ is the identity on
the output Hilbert space. For $k=1,\ldots,n$,
Eq.~(\ref{eq:pure-output}) implies
\begin{eqnarray}
P_\perp\Lambda_0(\Omega_k)P_\perp
=
\sum_\mu
\bigl(P_\perp K_\mu\Omega_k^{1/2}\bigr)
\bigl(P_\perp K_\mu\Omega_k^{1/2}\bigr)^\dagger
=
0 .
\label{eq:positive-sum}
\end{eqnarray}
Since all terms in Eq.~(\ref{eq:positive-sum}) are positive
semidefinite and their sum is zero, each term must vanish separately.
Consequently, $P_\perp K_\mu\Omega_k^{1/2}=0$ for every $\mu$. As the
range of $\Omega_k^{1/2}$ equals $\operatorname{supp}\Omega_k$, it
follows that
$K_\mu\bigl(\operatorname{supp}\Omega_k\bigr)
\subseteq\mathcal L_{\rm out}$ for every $\mu$. Given that every
$\tau_j$ is full rank, we have
$\operatorname{supp}\Omega_k=
\operatorname{span}\{|\Phi_d\rangle_k\}\otimes\mathcal H_{\bar k}$,
where
$\mathcal H_{\bar k}=\bigotimes_{j\neq k}
(\mathcal H_{A_j}\otimes\mathcal H_{B_j})$
is the Hilbert space of all pairs except $A_kB_k$. Accordingly, we
obtain
\begin{eqnarray}
K_\mu
\left(
|\Phi_d\rangle_k\otimes|\psi\rangle_{\bar k}
\right)
\in
\mathcal L_{\rm out},
\label{eq:support-constraint}
\end{eqnarray}
for every $\mu$, every $k=1,\ldots,n$, and every
$|\psi\rangle_{\bar k}\in\mathcal H_{\bar k}$.
In particular, the vector
$\ket{\Omega_0}:=\bigotimes_{j=1}^n|\Phi_d\rangle_j$ belongs to
$\operatorname{supp}\Omega_k$ for every $k=1,$$...$,$n$. We use
$\Omega_0$ to select a Kraus operator that acts nontrivially on this
common vector. Since $\Omega_0$ is pure, Eq.~(\ref{eq:pure-output}) gives
$\sum_\mu|v_\mu\rangle\langle v_\mu|
=p_0\Phi_d^{\rm out}$, where
$|v_\mu\rangle:=
K_\mu\ket{\Omega_0}$.
The condition $p_0>0$ ensures that at least one $|v_\mu\rangle$ is
nonzero. Moreover, the one-dimensional support of
$\Phi_d^{\rm out}$ implies that every $|v_\mu\rangle$ belongs to
$\mathcal L_{\rm out}$. We may therefore choose a product Kraus
operator, denoted by $K=M\otimes N$, such that
\begin{eqnarray}
K\ket{\Omega_0}
=
c|\Phi_d\rangle_{\rm out},
\qquad
c\neq0.
\label{eq:nonzero-anchor}
\end{eqnarray}
The selected $K$ therefore satisfies
Eq.~(\ref{eq:support-constraint}) for every $k=1,\ldots,n$.

We next
reduce $K$ to a product operator acting only on the first two pairs
while retaining the nonzero condition and the support constraints
corresponding to $k=1,2$. If $n=2$, we simply set $\tilde K=K$, and
no reduction is required. We therefore consider $n>2$. Let
$\mathcal R=\{3,\ldots,n\}$ denote the set of indices of the remaining
pairs. Using the local bases in which $|\Phi_d\rangle$ has the
standard form, for a multi-index
$\mathbf z=(z_3,\ldots,z_n)\in\{0,\ldots,d-1\}^{n-2}$, define
$|\mathbf z\rangle_{A_{\mathcal R}}
=\bigotimes_{j=3}^n|z_j\rangle_{A_j}$ and
$|\mathbf z\rangle_{B_{\mathcal R}}
=\bigotimes_{j=3}^n|z_j\rangle_{B_j}$.
The product of the maximally entangled states on the remaining pairs
can be expanded as
$\bigotimes_{j=3}^n|\Phi_d\rangle_j
=d^{-(n-2)/2}\sum_{\mathbf z}
|\mathbf z\rangle_{A_{\mathcal R}}
|\mathbf z\rangle_{B_{\mathcal R}}$.
For each $\mathbf z$, let us define the local operators
$M_{\mathbf z}:
\mathcal H_{A_1}\otimes\mathcal H_{A_2}\to\mathcal H_{A'}$
and
$N_{\mathbf z}:
\mathcal H_{B_1}\otimes\mathcal H_{B_2}\to\mathcal H_{B'}$
by
$M_{\mathbf z}|\xi\rangle
=M\left(|\xi\rangle\otimes
|\mathbf z\rangle_{A_{\mathcal R}}\right)$
and
$N_{\mathbf z}|\eta\rangle
=N\left(|\eta\rangle\otimes
|\mathbf z\rangle_{B_{\mathcal R}}\right)$.
Using Eq.~(\ref{eq:nonzero-anchor}), we obtain
$K\ket{\Omega_0}
=\frac{1}{\sqrt{d^{n-2}}}
\sum_{\mathbf z}(M_{\mathbf z}\otimes N_{\mathbf z})
\bigl(|\Phi_d\rangle_1\otimes|\Phi_d\rangle_2\bigr)\neq0$.
Hence there exists at least one multi-index $\mathbf z_0$ such that
\begin{eqnarray}
(M_{\mathbf z_0}\otimes N_{\mathbf z_0})
\bigl(
|\Phi_d\rangle_1\otimes|\Phi_d\rangle_2
\bigr)
\neq0 .
\label{eq:nonzero-slice}
\end{eqnarray}
Let us further define the corresponding product operator
$\tilde K:=\tilde M\otimes\tilde N$, with
$\tilde M:=M_{\mathbf z_0}$ and
$\tilde N:=N_{\mathbf z_0}$. For any
$|\phi\rangle_2\in
\mathcal H_{A_2}\otimes\mathcal H_{B_2}$, the vector
$|\Phi_d\rangle_1\otimes|\phi\rangle_2\otimes
|\mathbf z_0\rangle_{A_{\mathcal R}}
|\mathbf z_0\rangle_{B_{\mathcal R}}$
belongs to $\operatorname{supp}\Omega_1$, because every $\tau_j$ with
$j\neq1$ is full rank. Hence
Eq.~(\ref{eq:support-constraint}) implies
\begin{eqnarray}
\tilde K
\bigl(
|\Phi_d\rangle_1\otimes|\phi\rangle_2
\bigr)
\in
\mathcal L_{\rm out}.
\label{eq:nonzero-slice1}
\end{eqnarray}
Applying Eq.~(\ref{eq:support-constraint}) with $k=2$ yields
\begin{eqnarray}
\tilde K
\bigl(
|\phi\rangle_1\otimes|\Phi_d\rangle_2
\bigr)
\in
\mathcal L_{\rm out},
\label{eq:nonzero-slice2}
\end{eqnarray}
for every
$|\phi\rangle_1\in
\mathcal H_{A_1}\otimes\mathcal H_{B_1}$.

Third, we show that the nonzero condition in
Eq.~(\ref{eq:nonzero-slice}) and the support constraints in
Eqs.~(\ref{eq:nonzero-slice1}) and
(\ref{eq:nonzero-slice2}) are incompatible with the product form of
$\tilde K$.

We decompose $\tilde M$ and $\tilde N$ with respect to the local
subsystems $A_1$ and $B_1$ of the first pair as
$\tilde M=\sum_{i=0}^{d-1}
M_i(\langle i|_{A_1}\otimes I_{A_2})$ and
$\tilde N=\sum_{j=0}^{d-1}
N_j(\langle j|_{B_1}\otimes I_{B_2})$, where
$M_i:\mathcal H_{A_2}\to\mathcal H_{A'}$ and
$N_j:\mathcal H_{B_2}\to\mathcal H_{B'}$. Since these operators map
between $d$-dimensional spaces, we identify them with elements of
$M_d(\mathbb C)$ using the local bases in which $|\Phi_d\rangle$ has
the standard form. For any $X\in M_d(\mathbb C)$, define
$|\phi_X\rangle_j=(X\otimes I_{B_j})|\Phi_d\rangle_j$ for $j=1,2$.
As $X$ ranges over $M_d(\mathbb C)$, the vectors
$|\phi_X\rangle_j$ range over the entire space
$\mathcal H_{A_j}\otimes\mathcal H_{B_j}$. Using the
identity $(A\otimes B)|\Phi_d\rangle=(AB^T\otimes I_d)|\Phi_d\rangle$, where the superscript $T$ denotes transposition, we obtain
\begin{eqnarray}
\tilde K
\bigl(
|\Phi_d\rangle_1\otimes|\phi_X\rangle_2
\bigr)
=
\frac{1}{\sqrt d}
\left(
\sum_i M_iXN_i^T\otimes I_{B'}
\right)
|\Phi_d\rangle_{\rm out}.
\label{eq:first-block-action}
\end{eqnarray}
By Eq.~(\ref{eq:nonzero-slice1}), the vector in
Eq.~(\ref{eq:first-block-action}) belongs to
$\mathcal L_{\rm out}$ for every $X$. Since
$|\Phi_d\rangle_{\rm out}$ has full Schmidt rank, the map
$Y\mapsto(Y\otimes I_{B'})|\Phi_d\rangle_{\rm out}$ is injective.
Hence there exists a linear functional
$\alpha:M_d(\mathbb C)\to\mathbb C$ such that, for every
$X\in M_d(\mathbb C)$,
\begin{eqnarray}
\sum_i M_iXN_i^T
=
\alpha(X)I_{A'}.
\label{eq:first-matrix-constraint}
\end{eqnarray}
Similarly, expanding
$|\phi_X\rangle_1
=d^{-1/2}\sum_{r,s}X_{rs}|r\rangle_{A_1}|s\rangle_{B_1}$
gives
\begin{eqnarray}
\tilde K
\bigl(
|\phi_X\rangle_1\otimes|\Phi_d\rangle_2
\bigr)
=
\frac{1}{\sqrt d}
\left(
\sum_{r,s}X_{rs}M_rN_s^T
\otimes I_{B'}
\right)
|\Phi_d\rangle_{\rm out}.
\label{eq:second-block-action}
\end{eqnarray}
By Eq.~(\ref{eq:nonzero-slice2}), the vector in
Eq.~(\ref{eq:second-block-action}) belongs to
$\mathcal L_{\rm out}$ for every $X$. Hence there exists a linear
functional $\beta:M_d(\mathbb C)\to\mathbb C$ such that, for every
$X\in M_d(\mathbb C)$,
\begin{eqnarray}
\sum_{r,s}X_{rs}M_rN_s^T
=
\beta(X)I_{A'}.
\label{eq:second-matrix-constraint}
\end{eqnarray}
In particular, for $X=I_d$, Eqs.~(\ref{eq:first-block-action}) and
(\ref{eq:second-block-action}) both yield the same nonzero vector
$\tilde K\bigl(|\Phi_d\rangle_1\otimes|\Phi_d\rangle_2\bigr)$. Hence,
$\alpha(I_d)=\beta(I_d)\neq0$. Let $E_{rs}=|r\rangle\langle s|$ denote the standard matrix units.
Setting $X=E_{rs}$ in Eq.~(\ref{eq:second-matrix-constraint}) gives
\begin{eqnarray}
M_rN_s^T
=
c_{rs}I_{A'},
\label{eq:rank-one-constraint}
\end{eqnarray}
where $c_{rs}:=\beta(E_{rs})$.
Since $I_d=\sum_iE_{ii}$ and $\beta$ is linear, we have
$\sum_i c_{ii}=\beta(I_d)\neq0$. Hence there exists an index $t$ such
that $c_{tt}\neq0$. For this index, Eq.~(\ref{eq:rank-one-constraint}) gives
$M_tN_t^T=c_{tt}I_{A'}$. Since $c_{tt}\neq0$, both $M_t$ and $N_t$
are invertible. Setting $s=t$ in Eq.~(\ref{eq:rank-one-constraint})
and using the invertibility of $N_t^T$ gives
$M_r=\frac{c_{rt}}{c_{tt}}M_t$. Likewise, setting $r=t$ and using the
invertibility of $M_t$ gives
$N_s=\frac{c_{ts}}{c_{tt}}N_t$. Thus, there exist
$\uplambda_i,\mu_i\in\mathbb C$ such that
\begin{eqnarray}
M_i=\uplambda_iM_t
~~\text{and}~~
N_i=\mu_iN_t,
\label{eq:proportional-blocks}
\end{eqnarray}
for every $i$.
Substituting Eq.~(\ref{eq:proportional-blocks}) into
Eq.~(\ref{eq:first-matrix-constraint}), we have
\begin{eqnarray}
S M_tXN_t^T=\alpha(X)I_{A'}~~\text{with}~~
S:=\sum_i\uplambda_i\mu_i.
\label{eq:final-rigidity}
\end{eqnarray}
Taking $X=I_d$ and using
$M_tN_t^T=c_{tt}I_{A'}$, we obtain
$Sc_{tt}=\alpha(I_d)$. Since $\alpha(I_d)\neq0$ and
$c_{tt}\neq0$, it follows that $S\neq0$. Moreover, since
$N_t^T=c_{tt}M_t^{-1}$, Eq.~(\ref{eq:final-rigidity}) becomes
$Sc_{tt}M_tXM_t^{-1}=\alpha(X)I_{A'}$. Multiplying this equation
from the left by $M_t^{-1}$ and from the right by $M_t$, we obtain
$X=\frac{\alpha(X)}{Sc_{tt}}I_d$ for every
$X\in M_d(\mathbb C)$. However, choosing $X=|0\rangle\langle1|$ would require
$|0\rangle\langle1|$ to be proportional to $I_d$, which is impossible
for $d\geq2$. Hence no universal $n$-to-$1$ LOCC EPP exists for
$\mathcal S_d$.

It remains to consider the case $d_1\neq d_2$. Let
$d=\min\{d_1,d_2\}$, and suppose, to the contrary, that a universal
$n$-to-$1$ LOCC EPP for $\mathcal S_{d_1,d_2}$ exists. Repeat the
construction in Eq.~(\ref{eq:test-inputs}) with
$\tau_j=p\Phi_d^{(j)}+(1-p)I_{A_jB_j}/(d_1d_2)$, where
$1/(d+1)<p<1$. Each $\tau_j$ is full rank. It is also entangled for the chosen range
of $p$: indeed, its partial transpose $\tau_j^{T_{B_j}}$ has the
eigenvalue $(1-p)/(d_1d_2)-p/d<0$, where we have used
$d_1d_2\geq d^2$. Since
$\Phi_d^{(j)}$ is entangled as well, every $\Omega_k$ is an allowed
total input state. Since each $\Omega_k$ contains a maximally entangled state among its
$n$ input states, universality again gives
Eq.~(\ref{eq:pure-output}). Because every $\tau_j$ is full rank,
$\operatorname{supp}\Omega_k
=\operatorname{span}\{|\Phi_d\rangle_k\}\otimes\mathcal H_{\bar k}$.
Repeating the positivity and Kraus-selection arguments above, we
obtain a product Kraus operator $K=M\otimes N$ that satisfies
Eqs.~(\ref{eq:support-constraint}) and~(\ref{eq:nonzero-anchor}) for every $k$.
We next restrict $K$ to the $d$-dimensional
subspaces supporting $|\Phi_d\rangle$. In the local bases in which
$|\Phi_d\rangle$ has the standard form, define
$\mathcal K_{A_j}
=\operatorname{span}\{|r\rangle_{A_j}\}_{r=0}^{d-1}$ and
$\mathcal K_{B_j}
=\operatorname{span}\{|r\rangle_{B_j}\}_{r=0}^{d-1}$ for every
$j$, together with the corresponding output subspaces
$\mathcal K_{A'}$ and $\mathcal K_{B'}$.
Let $P_{A_j}$, $P_{B_j}$, $P_{A'}$, and $P_{B'}$ denote the
orthogonal projectors onto $\mathcal K_{A_j}$, $\mathcal K_{B_j}$,
$\mathcal K_{A'}$, and $\mathcal K_{B'}$, respectively, and set
$P_A^{\rm in}:=\bigotimes_{j=1}^nP_{A_j}$ and
$P_B^{\rm in}:=\bigotimes_{j=1}^nP_{B_j}$. Define
$\widehat M:=P_{A'}MP_A^{\rm in}$ and
$\widehat N:=P_{B'}NP_B^{\rm in}$, and
$\widehat K:=\widehat M\otimes\widehat N$.
The input projections $P_A^{\rm in}$ and $P_B^{\rm in}$ act as the identity on
$\bigotimes_{j=1}^n\mathcal K_{A_j}$ and
$\bigotimes_{j=1}^n\mathcal K_{B_j}$, while the output projection
$P_{A'}\otimes P_{B'}$ acts as the identity on
$\mathcal L_{\rm out}
=\operatorname{span}\{|\Phi_d\rangle_{\rm out}\}
\subseteq\mathcal K_{A'}\otimes\mathcal K_{B'}$.
Hence $\widehat K$ remains nonzero on
$\bigotimes_{j=1}^n|\Phi_d\rangle_j$ and satisfies the restrictions of
Eq.~(\ref{eq:support-constraint}) to the chosen input subspaces.
Applying to $\widehat K$ the same reduction to the first two pairs
used above yields a nonzero product operator
$\tilde K=\tilde M\otimes\tilde N$ satisfying the
restrictions of Eqs.~(\ref{eq:nonzero-slice1}) and
(\ref{eq:nonzero-slice2}) to the chosen $d$-dimensional input and
output subspaces. Since all these subspaces are $d$-dimensional,
$\tilde M$ and $\tilde N$ may be decomposed as in the
$d_1=d_2$ case, yielding Eqs.~(\ref{eq:first-matrix-constraint})
and (\ref{eq:second-matrix-constraint}). Therefore, the matrix argument developed above for the case $d_1=d_2$ applies without modification and yields the same contradiction.
 \hfill$\blacksquare$

Having established that no universal EPP exists in the bipartite
setting, we next turn to the multipartite case.

\textbf{Theorem 2.} For any number of parties $m\geq3$, any fixed finite local dimensions
$\boldsymbol d=(d_X)_{X\in\mathcal P}$ with $d_X\geq2$, and any $n\geq2$, no $n$-to-$1$ LOCC EPP is universal for $\mathcal G_{\boldsymbol d}$.

\emph{Proof sketch.} For each $j=1,\ldots,n$, let
$\Psi_{\boldsymbol d}^{(j)}$ denote the generalized GHZ
state on
$\bigotimes_{X\in\mathcal P}\mathcal H_{X_j}$, and define
$\tau_j=p\Psi_{\boldsymbol d}^{(j)}+(1-p)I_j/D$, where
$(D-d)/[d(D-1)]<p<1$. Each $\tau_j$ is full rank and satisfies
$F(\tau_j)>1/d$. Since the fidelity of every biseparable state with $\Psi_{\boldsymbol d}^{(j)}$ is at most $1/d$, each $\tau_j$ is
genuinely multipartite entangled for the chosen range of $p$. Consider the $n+1$ inputs
$\Theta_0:=\bigotimes_{j=1}^n\Psi_{\boldsymbol d}^{(j)}$ and
$\Theta_k:=\Psi_{\boldsymbol d}^{(k)}
\otimes\bigotimes_{j\neq k}\tau_j$ for $k=1,\ldots,n$.
Each $\Theta_k$ contains a $\Psi_{\boldsymbol d}$, so universality
requires its normalized successful output to be the
target state $\Psi_{\boldsymbol d}$. The positivity and Kraus-selection arguments of
Theorem~1 then yield the corresponding support constraints and a
product Kraus operator that is nonzero on $\Theta_0$. Across any bipartition of the parties, $\Psi_{\boldsymbol d}$ is
maximally entangled with Schmidt rank $d$, while every multipartite
product Kraus operator remains product across that bipartition.
Treating the two sides of the bipartition as two effective parties,
the support constraints therefore reduce to those in the bipartite
proof, giving the same contradiction. The complete proof is provided
in the Supplemental Material~\cite{SM}.

We note that the proofs of Theorems~1 and 2 rely only on the product
form of the Kraus operators. Since separable
operations~\cite{Rains1999} also admit such product Kraus representations, both no-go theorems remain valid under separable
operations. We next consider two further extensions of these results.

First, we show that the no-go results in Theorems~1 and 2 persist after restricting the
allowed inputs to states above any prescribed fidelity threshold. Fix
$0<\epsilon<1$. In the bipartite case, define
$\mathcal S_{d_1,d_2}^{(\epsilon)}
:=\{\rho\in\mathcal S_{d_1,d_2}:F(\rho)>1-\epsilon\}$,
and let $d=\min\{d_1,d_2\}$. The states used in the proof of
Theorem~1 satisfy $F(\Phi_d)=1$ and
$F(\tau_j)=p+(1-p)/(d_1d_2)$. For
$1/(d+1)<p<1$, every $\tau_j$ is full rank and entangled, and
$F(\tau_j)\to1$ as $p\to1$. We may therefore choose $p$ within this
range so that $F(\tau_j)>1-\epsilon$. All states used in the
construction then belong to
$\mathcal S_{d_1,d_2}^{(\epsilon)}$, while the full-rank property of
the states $\tau_j$, and hence the support constraints derived in
Theorem~1, remain unchanged. The same contradiction therefore shows
that no $n$-to-$1$ LOCC EPP is universal for
$\mathcal S_{d_1,d_2}^{(\epsilon)}$ for any $n\geq2$. The multipartite case is analogous. Define
$\mathcal G_{\boldsymbol d}^{(\epsilon)}
:=\{\rho\in\mathcal G_{\boldsymbol d}:F(\rho)>1-\epsilon\}$,
and let $d=\min_{X\in\mathcal P}d_X$ and
$D=\prod_{X\in\mathcal P}d_X$. In the multipartite construction,
$F(\Psi_{\boldsymbol d})=1$ and
$F(\tau_j)=p+(1-p)/D$. For
$(D-d)/[d(D-1)]<p<1$, every $\tau_j$ is full rank and genuinely
multipartite entangled, and $F(\tau_j)\to1$ as $p\to1$.
Choosing $p<1$ sufficiently close to one within this range places all
states used in the construction in
$\mathcal G_{\boldsymbol d}^{(\epsilon)}$ without changing the
support constraints derived in Theorem~2. The same contradiction
therefore applies. These observations give the following corollary.

\textbf{Corollary 1.} For every $0<\epsilon<1$ and $n\geq2$, no
$n$-to-$1$ LOCC EPP is universal for any
$\mathcal S_{d_1,d_2}^{(\epsilon)}$ with $d_1,d_2\geq2$ or any
$\mathcal G_{\boldsymbol d}^{(\epsilon)}$ with
$\boldsymbol d=(d_X)_{X\in\mathcal P}$ finite,
$|\mathcal P|\geq3$, and $d_X\geq2$.

Second, we show that the no-go results in Theorems~1 and 2
persist even when the protocol is assisted by a finite-dimensional pure catalyst chosen independently of the input states
~\cite{Jonathan1999,Datta2023,LipkaBartosik2024}. Let $|\chi\rangle_C$ be a catalyst shared among the parties and write
$\chi:=|\chi\rangle_C\langle\chi|$. An $n$-to-$1$ catalytic EPP with
successful branch $\Gamma_0$ is called universal if, for every allowed total input state
$\rho^{\mathrm{in}}=\bigotimes_{j=1}^n\rho_j$, there exist a success
probability $q_{\rho^{\mathrm{in}}}>0$ and a normalized output state
$\omega_{\rho^{\mathrm{in}}}$ such that
$\Gamma_0(\rho^{\mathrm{in}}\otimes\chi)=q_{\rho^{\mathrm{in}}}\omega_{\rho^{\mathrm{in}}}\otimes\chi$
and
$F(\omega_{\rho^{\mathrm{in}}})
\geq\max_{1\leq j\leq n}F(\rho_j)$. The following theorem extends Theorems~1 and 2 to catalytic EPPs. Its proof is given in the
Supplemental Material~\cite{SM}.

\textbf{Theorem 3.} Universal EPPs remain impossible under LOCC on
the full bipartite and multipartite entangled-state sets
considered in Theorems~1 and 2, even with the assistance of pure catalyst.

Before concluding, we note that the no-go results have an analogue in
coherence purification \cite{Winter2016,Fang2018,Lami2019,Liu2019}. Under strictly
incoherent operations~\cite{Baumgratz2014,Winter2016,Yadin2016,Streltsov2017},
we establish that no
universal $n$-to-$1$ coherence-purification protocol exists for any
$n\geq2$, even for inputs restricted to fidelities above an arbitrarily high threshold below one.
The definitions and proof are given in the Supplemental
Material~\cite{SM}.

\section{Conclusions}
We have established a general no-go theorem for finite-input universal
EPPs: for every $n\geq2$ and every fixed choice of finite local
dimensions, no $n$-to-$1$ LOCC protocol is universal for either the
full set of bipartite entangled states or the full set of
multipartite entangled states. This settles the higher-dimensional
bipartite and multipartite cases left open in
Ref.~\cite{Zang2025}. Taken together, the extensions established above
and the analogous coherence no-go result show that the obstruction is
neither specific to LOCC nor unique to EPP.

Importantly, these no-go results do not end the search for universal
purification, but redirect it toward physically relevant input
classes. In a given experiment, the exact input states may be unknown,
while the preparation conditions and dominant error processes may
still be partially characterized or constrained \cite{Breuer2002}. A central question is therefore how to use such
information to identify the broadest admissible input class for which
a single EPP can be universal. Addressing this question can guide the
design of EPPs for quantum computation, quantum communication, and
quantum cryptography, while sharpening our understanding of the limits of
entanglement purification.

\section*{Acknowledgements}
This work was supported by the National Natural Science Foundation of
China (Grant No.~12405019) and the Natural Science Foundation of
Shandong Province (Grant No.~ZR2025LLZ002).

\section*{Supplemental Material}

In this Supplemental Material, we first provide the complete proof of
Theorem~2 in the main text. We then prove Theorem~3, which extends Theorems~1 and~2 to catalytic
EPPs.
Finally, we give the definitions and proof of the no-go theorem for
universal coherence purification under strictly incoherent operations,
including the case of input sets whose minimum fidelity with the
maximally coherent state can be chosen arbitrarily close to one.

Throughout, an identity operator acting on a Hilbert space is labeled
by the corresponding space or subsystem. In particular,
$I_{\mathcal K}$ denotes the identity operator on $\mathcal K$. In
matrix representations, $I_r$ denotes the $r\times r$ identity
matrix.

\section*{SI.~~Complete proof of Theorem 2}

\textbf{Theorem 2.} For any number of parties $m\geq3$, any fixed finite local dimensions
$\boldsymbol d=(d_X)_{X\in\mathcal P}$ with $d_X\geq2$, and any $n\geq2$, no $n$-to-$1$ LOCC EPP is universal for $\mathcal G_{\boldsymbol d}$.

\textit{Proof.} We argue by contradiction and suppose that a universal
$n$-to-$1$ LOCC EPP for $\mathcal G_{\boldsymbol d}$ exists.

First, we construct total input states for which universality requires
the normalized output conditioned on success to be the generalized
GHZ state.

Recall that $d=\min_{X\in\mathcal P}d_X$ and
$D=\prod_{X\in\mathcal P}d_X$. For each $j=1,\ldots,n$, let
$\Psi_{\boldsymbol d}^{(j)}:=|\Psi_{\boldsymbol d}\rangle_j
\langle\Psi_{\boldsymbol d}|$ denote the generalized GHZ
state on $\bigotimes_{X\in\mathcal P}\mathcal H_{X_j}$ and define
$|\Psi_{\boldsymbol d}\rangle_j=d^{-1/2}\sum_{r=0}^{d-1}
\bigotimes_{X\in\mathcal P}|r\rangle_{X_j}$.
Define
$\tau_j=p\Psi_{\boldsymbol d}^{(j)}+(1-p)I_j/D$, where $I_j$ is the
identity on $\bigotimes_{X\in\mathcal P}\mathcal H_{X_j}$ and
$(D-d)/[d(D-1)]<p<1$. Each $\tau_j$ is full rank and satisfies
$F(\tau_j)=p+(1-p)/D>1/d$. Across every bipartition of the parties,
$|\Psi_{\boldsymbol d}\rangle_j$ has $d$ equal Schmidt coefficients.
Thus, the fidelity of any pure state that is product across that
bipartition with $\Psi_{\boldsymbol d}^{(j)}$ is at most $1/d$.
Since every biseparable state is a convex combination of pure states
that are product across possibly different bipartitions, its fidelity with $\Psi_{\boldsymbol d}^{(j)}$ is also at most
$1/d$. It follows that every $\tau_j$ is genuinely multipartite entangled
and hence belongs to $\mathcal G_{\boldsymbol d}$.
Let us consider the following $n+1$ total input states:
\begin{eqnarray}
\Theta_0
=
\bigotimes_{j=1}^{n}\Psi_{\boldsymbol d}^{(j)},
~~
\Theta_k
=
\Psi_{\boldsymbol d}^{(k)}
\otimes
\bigotimes_{j\neq k}\tau_j,
~~
k=1,\ldots,n .
\label{eq:sm-multi-inputs}
\end{eqnarray}
Every input state in each $\Theta_k$ belongs to
$\mathcal G_{\boldsymbol d}$, so every $\Theta_k$ is an allowed total
input state. Let $|\Psi_{\boldsymbol d}\rangle_{\rm out}$ denote the
 target state on
$\bigotimes_{X\in\mathcal P}\mathcal H_{X'}$, with
$\Psi_{\boldsymbol d}^{\rm out}:=
|\Psi_{\boldsymbol d}\rangle_{\rm out}
\langle\Psi_{\boldsymbol d}|$. Since each of $\Theta_0,\ldots,\Theta_n$ contains at least one input
state equal to the corresponding target state, the quantity
$\max_{1\leq j\leq n}F(\rho_j)$ in the universality condition equals
one. Therefore, the normalized successful output must have unit
fidelity with $|\Psi_{\boldsymbol d}\rangle_{\rm out}$. Equivalently,
\begin{eqnarray}
\Lambda_0(\Theta_k)
=
p_k\Psi_{\boldsymbol d}^{\rm out},
\qquad
k=0,\ldots,n,
\label{eq:sm-multi-pure-output}
\end{eqnarray}
where $p_k:=\operatorname{Tr}[\Lambda_0(\Theta_k)]>0$. Since $\Lambda_0$ is the successful branch of an LOCC protocol, it
admits a product Kraus representation
$\Lambda_0(\rho)=\sum_\mu K_\mu\rho K_\mu^\dagger$, where
$K_\mu=\bigotimes_{X\in\mathcal P}M_{\mu,X}$, with
$M_{\mu,X}:\bigotimes_{j=1}^{n}\mathcal H_{X_j}
\to\mathcal H_{X'}$.

Second, we derive support constraints for every product Kraus operator
in the successful branch and then select one satisfying a nonzero
condition.

Let $\mathcal L_{\rm out}$ be the subspace spanned by
$|\Psi_{\boldsymbol d}\rangle_{\rm out}$, and let $P_\perp$ be the
orthogonal projector onto $\mathcal L_{\rm out}^{\perp}$. For every
$k=1,\ldots,n$, Eq.~(\ref{eq:sm-multi-pure-output}) implies
\begin{eqnarray}
P_\perp\Lambda_0(\Theta_k)P_\perp
=
\sum_\mu
\bigl(P_\perp K_\mu\Theta_k^{1/2}\bigr)
\bigl(P_\perp K_\mu\Theta_k^{1/2}\bigr)^\dagger
=
0 .
\label{eq:sm-multi-positive-sum}
\end{eqnarray}
Since all terms in Eq.~(\ref{eq:sm-multi-positive-sum}) are positive
semidefinite and their sum is zero, each term must vanish separately.
Consequently, $P_\perp K_\mu\Theta_k^{1/2}=0$ for every $\mu$. As the
range of $\Theta_k^{1/2}$ equals $\operatorname{supp}\Theta_k$, it
follows that
$K_\mu\bigl(\operatorname{supp}\Theta_k\bigr)
\subseteq\mathcal L_{\rm out}$ for every $\mu$. Given that every
$\tau_j$ is full rank, we have
$\operatorname{supp}\Theta_k=
\operatorname{span}\{|\Psi_{\boldsymbol d}\rangle_k\}
\otimes\mathcal H_{\bar k}$, where
$\mathcal H_{\bar k}=\bigotimes_{j\neq k}
\bigotimes_{X\in\mathcal P}\mathcal H_{X_j}$ is
the Hilbert space of all input factors except the kth. Accordingly, we obtain
\begin{eqnarray}
K_\mu
\left(
|\Psi_{\boldsymbol d}\rangle_k
\otimes|\psi\rangle_{\bar k}
\right)
\in
\mathcal L_{\rm out},
\label{eq:sm-multi-support}
\end{eqnarray}
for every $\mu$, every $k=1,\ldots,n$, and every
$|\psi\rangle_{\bar k}\in\mathcal H_{\bar k}$.
In particular,
$\bigotimes_{j=1}^{n}|\Psi_{\boldsymbol d}\rangle_j$ belongs to
$\operatorname{supp}\Theta_k$ for every $k=1,\ldots,n$. We use
$\Theta_0$ to select a Kraus operator that acts nontrivially on this
common state. Since $\Theta_0$ is pure,
Eq.~(\ref{eq:sm-multi-pure-output}) gives
$\sum_\mu|v_\mu\rangle\langle v_\mu|
=p_0\Psi_{\boldsymbol d}^{\rm out}$, where
$|v_\mu\rangle:=K_\mu
\bigotimes_{j=1}^{n}|\Psi_{\boldsymbol d}\rangle_j$.
The condition $p_0>0$ ensures that at least one $|v_\mu\rangle$ is
nonzero. Moreover, the one-dimensional support of
$\Psi_{\boldsymbol d}^{\rm out}$ implies that every $|v_\mu\rangle$
belongs to $\mathcal L_{\rm out}$. We may therefore choose a product
Kraus operator, denoted by
$K=\bigotimes_{X\in\mathcal P}M_X$, such that
\begin{eqnarray}
K
\bigotimes_{j=1}^{n}|\Psi_{\boldsymbol d}\rangle_j
=
c|\Psi_{\boldsymbol d}\rangle_{\rm out},
\qquad
c\neq0 .
\label{eq:sm-multi-anchor}
\end{eqnarray}
The selected $K$ therefore satisfies
Eq.~(\ref{eq:sm-multi-support}) for every $k=1,\ldots,n$.

We next reduce $K$ to a product operator acting only on the first two
input factors while retaining the nonzero condition and the support
constraints corresponding to $k=1,2$. If $n=2$, we set
$\tilde K=K$ and $\tilde M_X=M_X$ for every $X\in\mathcal P$, and no
reduction is required. We therefore consider $n>2$. Let $\mathcal R=\{3,\ldots,n\}$ denote the index set of the remaining
input factors. For a multi-index
$\mathbf z=(z_3,\ldots,z_n)\in\{0,\ldots,d-1\}^{n-2}$, define
$|\mathbf z\rangle_{X_{\mathcal R}}
=\bigotimes_{j=3}^{n}|z_j\rangle_{X_j}$ for every
$X\in\mathcal P$. Thus, $\bigotimes_{j=3}^{n}|\Psi_{\boldsymbol d}\rangle_j$ can be expanded as
\begin{eqnarray}
\bigotimes_{j=3}^{n}|\Psi_{\boldsymbol d}\rangle_j
=
\frac{1}{\sqrt{d^{n-2}}}
\sum_{\mathbf z}
\bigotimes_{X\in\mathcal P}
|\mathbf z\rangle_{X_{\mathcal R}}.
\label{eq:sm-multi-expansion}
\end{eqnarray}
For each $\mathbf z$, define the local operators
$M_{X,\mathbf z}:\mathcal H_{X_1}\otimes\mathcal H_{X_2}
\to\mathcal H_{X'}$ by
$M_{X,\mathbf z}|\xi\rangle
=M_X\bigl(|\xi\rangle\otimes
|\mathbf z\rangle_{X_{\mathcal R}}\bigr)$.
Using Eqs.~(\ref{eq:sm-multi-anchor}) and
(\ref{eq:sm-multi-expansion}), we obtain
\begin{eqnarray}
K\bigotimes_{j=1}^{n}|\Psi_{\boldsymbol d}\rangle_j
=
\frac{1}{\sqrt{d^{n-2}}}
\sum_{\mathbf z}
\left(
\bigotimes_{X\in\mathcal P}M_{X,\mathbf z}
\right)
\left(
|\Psi_{\boldsymbol d}\rangle_1
\otimes|\Psi_{\boldsymbol d}\rangle_2
\right)
\neq0.\nonumber\\
\label{eq:sm-multi-slice-sum}
\end{eqnarray}
Hence there exists at least one multi-index $\mathbf z_0$ such that
the product operator
$\tilde K:=\bigotimes_{X\in\mathcal P}\tilde M_X$, with
$\tilde M_X:=M_{X,\mathbf z_0}$, satisfies
\begin{eqnarray}
\tilde K
\left(
|\Psi_{\boldsymbol d}\rangle_1
\otimes|\Psi_{\boldsymbol d}\rangle_2
\right)
\neq0 .
\label{eq:sm-multi-nonzero}
\end{eqnarray}
For $n>2$ and any
$|\phi\rangle_2\in
\bigotimes_{X\in\mathcal P}\mathcal H_{X_2}$, the vector
$|\Psi_{\boldsymbol d}\rangle_1\otimes|\phi\rangle_2
\otimes\bigotimes_{X\in\mathcal P}
|\mathbf z_0\rangle_{X_{\mathcal R}}$
belongs to $\operatorname{supp}\Theta_1$, because every $\tau_j$ with
$j\neq1$ is full rank. Hence Eq.~(\ref{eq:sm-multi-support}) implies
\begin{eqnarray}
\tilde K
\left(
|\Psi_{\boldsymbol d}\rangle_1
\otimes|\phi\rangle_2
\right)
\in
\mathcal L_{\rm out}.
\label{eq:sm-multi-slice1}
\end{eqnarray}
Applying Eq.~(\ref{eq:sm-multi-support}) with $k=2$ yields
\begin{eqnarray}
\tilde K
\left(
|\phi\rangle_1
\otimes|\Psi_{\boldsymbol d}\rangle_2
\right)
\in
\mathcal L_{\rm out},
\label{eq:sm-multi-slice2}
\end{eqnarray}
for every
$|\phi\rangle_1\in
\bigotimes_{X\in\mathcal P}\mathcal H_{X_1}$.

Third, we show that the nonzero property of $\tilde K$ established
in Eq. (\ref{eq:sm-multi-nonzero}) and the two support constraints in
Eqs.~(\ref{eq:sm-multi-slice1}) and
(\ref{eq:sm-multi-slice2}) reduce to the incompatible bipartite
conditions derived in the proof of Theorem~1.

Consider the bipartition $A\,|\,\bar A$, where
$\bar A:=\mathcal P\setminus\{A\}$. For $j=1,2$, define
$\mathcal U_j
=\operatorname{span}\{|r\rangle_{A_j}:0\leq r\leq d-1\}$ and
$\mathcal V_j
=\operatorname{span}\{
\bigotimes_{X\in\bar A}|r\rangle_{X_j}:0\leq r\leq d-1\}$,
and define the corresponding output subspaces
$\mathcal U_{\rm out}$ and $\mathcal V_{\rm out}$ analogously. Let
$P_{\mathcal U_j}$, $P_{\mathcal V_j}$,
$P_{\mathcal U_{\rm out}}$, and $P_{\mathcal V_{\rm out}}$ denote
the orthogonal projectors onto these subspaces.
After grouping the tensor factors according to the bipartition
$A\,|\,\bar A$, write
$\tilde K=\tilde M_A\otimes\tilde N_{\bar A}$, where
$\tilde N_{\bar A}:=
\bigotimes_{X\in\bar A}\tilde M_X$. Define a product
operator $\widehat K:=\widehat M\otimes\widehat N$, where
\begin{eqnarray}
\widehat M
&:=&
P_{\mathcal U_{\rm out}}
\tilde M_A
\left(
P_{\mathcal U_1}\otimes P_{\mathcal U_2}
\right),
\nonumber\\
\widehat N
&:=&
P_{\mathcal V_{\rm out}}
\tilde N_{\bar A}
\left(
P_{\mathcal V_1}\otimes P_{\mathcal V_2}
\right).
\label{eq:sm-multi-restricted-K}
\end{eqnarray}
For each $j=1,2$, identify
$\bigotimes_{X\in\bar A}|r\rangle_{X_j}$ with a basis vector
$|r\rangle_{\bar A_j}$ of $\mathcal V_j$. Under this identification,
\begin{eqnarray}
|\Psi_{\boldsymbol d}\rangle_j
=
\frac{1}{\sqrt d}
\sum_{r=0}^{d-1}
|r\rangle_{A_j}|r\rangle_{\bar A_j}
=:
|\Phi_d\rangle_j .
\label{eq:sm-multi-logical-bell}
\end{eqnarray}
We use the same identification on the output Hilbert space and denote
the target state by $|\Phi_d\rangle_{\rm out}$. All vectors of the forms appearing on the left-hand sides of
Eqs.~(\ref{eq:sm-multi-slice1}) and
(\ref{eq:sm-multi-slice2}) belong to
$\mathcal L_{\rm out}$ and are therefore proportional to
$|\Psi_{\boldsymbol d}\rangle_{\rm out}$, which lies in
$\mathcal U_{\rm out}\otimes\mathcal V_{\rm out}$.
Consequently, the projection
$P_{\mathcal U_{\rm out}}\otimes P_{\mathcal V_{\rm out}}$
leaves these vectors unchanged. It follows that
\begin{eqnarray}
\widehat K
\left(
|\Phi_d\rangle_1\otimes|\phi\rangle_2
\right)
&\in&
\operatorname{span}\{|\Phi_d\rangle_{\rm out}\},
\nonumber\\
\widehat K
\left(
|\phi\rangle_1\otimes|\Phi_d\rangle_2
\right)
&\in&
\operatorname{span}\{|\Phi_d\rangle_{\rm out}\},
\label{eq:sm-multi-bipartite-support}
\end{eqnarray}
for every
$|\phi\rangle_j\in\mathcal U_j\otimes\mathcal V_j$, $j=1,2$. Likewise, Eq.~(\ref{eq:sm-multi-nonzero}) gives
\begin{eqnarray}
\widehat K
\left(
|\Phi_d\rangle_1\otimes|\Phi_d\rangle_2
\right)
\neq0 .
\label{eq:sm-multi-bipartite-nonzero}
\end{eqnarray}
Since the spaces $\mathcal U_1$, $\mathcal U_2$,
$\mathcal U_{\rm out}$, $\mathcal V_1$, $\mathcal V_2$, and
$\mathcal V_{\rm out}$ are all $d$-dimensional,
Eqs.~(\ref{eq:sm-multi-bipartite-support}) and
(\ref{eq:sm-multi-bipartite-nonzero}) are precisely the two support
constraints and the nonzero condition used in the matrix argument for
the case $d_1=d_2$ in the proof of Theorem~1.
 That argument shows that no nonzero bipartite
product operator $\widehat K$ can satisfy these conditions for
$d\geq2$. This contradiction means that no universal $n$-to-$1$
LOCC EPP exists for $\mathcal G_{\boldsymbol d}$ for any fixed finite
local dimensions $\boldsymbol d=(d_X)_{X\in\mathcal P}$.
\hfill$\blacksquare$

\section*{SII.~~Exact pure-state catalytic assistance}

Previous studies have shown that catalytic assistance can enable state
transformations that would otherwise be impossible under
LOCC~\cite{Jonathan1999,Datta2023,LipkaBartosik2024}. In this section,
we investigate whether universal entanglement purification can be
achieved with the assistance of a finite-dimensional pure
catalyst. Let $|\chi\rangle_C\in\mathcal H_C$ be the pure catalyst
state shared among the parties and chosen independently of the input
states, where
$\mathcal H_C:=\bigotimes_{X\in\mathcal P}\mathcal H_{C_X}$ and
$\mathcal H_{C_X}$ is a finite-dimensional Hilbert space associated
with party $X$. We write
$\chi:=|\chi\rangle_C\langle\chi|$.

The successful branch $\Gamma_0$ of an $n$-to-$1$ EPP is universal
with the assistance of the catalyst $\chi$ if, for every allowed
total input state
$\rho^{\mathrm{in}}=\bigotimes_{j=1}^n\rho_j$, it satisfies
\begin{eqnarray}
\Gamma_0(\rho^{\mathrm{in}}\otimes\chi)
=
q_{\rho^{\mathrm{in}}}
\omega_{\rho^{\mathrm{in}}}\otimes\chi
\label{eq:sm-cat-return}
\end{eqnarray}
and
$F(\omega_{\rho^{\mathrm{in}}})
\geq\max_{1\leq j\leq n}F(\rho_j)$,
where $q_{\rho^{\mathrm{in}}}>0$ is the success probability,
$\omega_{\rho^{\mathrm{in}}}$ is the normalized output state, and the
tensor-product form on the right-hand side expresses that the
catalyst is returned exactly and without correlations with the
output.

With the above definitions in place, we begin the proof of
Theorem~3 in the main text by establishing the following lemma.

\textbf{Lemma.}
Suppose that the local Hilbert spaces associated
with $A_1,A_2,A',B_1,B_2$, and $B'$ are all $d(\geq2)$-dimensional. Let
$|\Phi_d\rangle_j$, $j=1,2$, and
$|\Phi_d\rangle_{\rm out}$ denote the corresponding maximally
entangled states on the two input pairs and the output pair,
respectively. Let
$|\chi\rangle_{C_AC_B}\in
\mathcal H_{C_A}\otimes\mathcal H_{C_B}$ be any pure state, where
$\mathcal H_{C_A}$ and $\mathcal H_{C_B}$ are finite-dimensional, and
define
$\mathcal L_{\rm cat}:=
\operatorname{span}\{
|\Phi_d\rangle_{\rm out}\otimes|\chi\rangle\}$.
There is no product operator
$\tilde K=\tilde M\otimes\tilde N$, with
$\tilde M:\mathcal H_{A_1}\otimes\mathcal H_{A_2}\otimes\mathcal H_{C_A}\to\mathcal H_{A'}\otimes\mathcal H_{C_A},$ and
$\tilde N:
\mathcal H_{B_1}\otimes\mathcal H_{B_2}\otimes\mathcal H_{C_B}
\to
\mathcal H_{B'}\otimes\mathcal H_{C_B},$
that simultaneously satisfies the nonzero condition
\begin{eqnarray}
\tilde K
\left(
|\Phi_d\rangle_1\otimes|\Phi_d\rangle_2
\otimes|\chi\rangle
\right)
\neq0
\label{eq:sm-cat-nonzero}
\end{eqnarray}
and the two support constraints
\begin{eqnarray}
\tilde K
\left(
|\Phi_d\rangle_1\otimes|\phi\rangle_2
\otimes|\chi\rangle
\right)
&\in&
\mathcal L_{\rm cat},
\nonumber\\
\tilde K
\left(
|\phi\rangle_1\otimes|\Phi_d\rangle_2
\otimes|\chi\rangle
\right)
&\in&
\mathcal L_{\rm cat}
\label{eq:sm-cat-two-support}
\end{eqnarray}
for every
$|\phi\rangle_j\in
\mathcal H_{A_j}\otimes\mathcal H_{B_j}$, $j=1,2$.

\textit{Proof.}~First, we restrict the catalyst
state to its Schmidt-support subspaces and translate the two
support constraints into matrix equations.

Let the Schmidt decomposition of the catalyst be
$|\chi\rangle=\sum_{a=1}^{R_\chi}\sqrt{\uplambda_a}
|a\rangle_{C_A}|a\rangle_{C_B}$, where $R_\chi$ is its Schmidt rank. Define
$\mathcal S_A^\chi:=
\operatorname{span}\{|a\rangle_{C_A}:1\leq a\leq R_\chi\}$
and
$\mathcal S_B^\chi:=
\operatorname{span}\{|a\rangle_{C_B}:1\leq a\leq R_\chi\}$,
and let $P_A^\chi$ and $P_B^\chi$ denote the orthogonal projectors
onto $\mathcal S_A^\chi$ and $\mathcal S_B^\chi$, respectively. Define
\begin{align}
\overline M
&:=
(I_{A'}\otimes P_A^\chi)
\tilde M
(I_{A_1}\otimes I_{A_2}\otimes P_A^\chi),
\nonumber\\
\overline N
&:=
(I_{B'}\otimes P_B^\chi)
\tilde N
(I_{B_1}\otimes I_{B_2}\otimes P_B^\chi).
\label{eq:sm-cat-support-restriction}
\end{align}
and set $\overline K:=\overline M\otimes\overline N$. The input
projections $I_{A_1}\otimes I_{A_2}\otimes P_A^\chi$ and $I_{B_1}\otimes I_{B_2}\otimes P_B^\chi$ leave the catalyst unchanged because
$(P_A^\chi\otimes P_B^\chi)|\chi\rangle=|\chi\rangle$.
Moreover, taking
$|\phi\rangle_2=|\Phi_d\rangle_2$ in the first support condition of
Eq.~(\ref{eq:sm-cat-two-support}) and using
Eq.~(\ref{eq:sm-cat-nonzero}), we find that the vector in
Eq.~(\ref{eq:sm-cat-nonzero}) is a nonzero element of
$\mathcal L_{\rm cat}$.
The output projections $I_{A'}\otimes P_A^\chi$ and $I_{B'}\otimes P_B^\chi$ therefore leave unchanged all vectors on the
left-hand sides of Eqs.~(\ref{eq:sm-cat-nonzero}) and
(\ref{eq:sm-cat-two-support}). Consequently, Eqs.~(\ref{eq:sm-cat-nonzero}) and
(\ref{eq:sm-cat-two-support}) remain valid with $\tilde K$ replaced
by $\overline K$. We may therefore replace $\mathcal H_{C_A}$ by
$\mathcal S_A^\chi$ and $\mathcal H_{C_B}$ by
$\mathcal S_B^\chi$. For notational simplicity, we relabel $\overline M$ and
$\overline N$ as $\tilde M$ and $\tilde N$, respectively.
Define $D_\chi:=\operatorname{diag}
(\sqrt{\uplambda_1},\ldots,\sqrt{\uplambda_{R_\chi}})$
and $Q_\chi:=I_d\otimes D_\chi$. Since $\uplambda_a>0$ for all
$a=1,\ldots,R_\chi$, both $D_\chi$ and $Q_\chi$ are invertible.
Decompose $\tilde M$ and $\tilde N$ with respect to
$A_1$ and $B_1$ as
$\tilde M=\sum_{i=0}^{d-1}M_i
(\langle i|_{A_1}\otimes I_{A_2}\otimes I_{C_A})$
and
$\tilde N=\sum_{j=0}^{d-1}N_j
(\langle j|_{B_1}\otimes I_{B_2}\otimes I_{C_B})$.
Here, $I_{C_A}$ and $I_{C_B}$ denote the identity operators on
$\mathcal S_A^\chi$ and $\mathcal S_B^\chi$, respectively. Hence every $M_i$ and $N_j$ is represented
by a $dR_\chi\times dR_\chi$ matrix. For a bipartite vector
$|v\rangle=\sum_{a,b=1}^{dR_\chi}
V_{ab}|a\rangle_L|b\rangle_R$, define its coefficient matrix as
$V:=(V_{ab})\in M_{dR_\chi}(\mathbb C)$.
If local operators $A$ and $B$ act on the two subsystems,
respectively, then the coefficient matrix of
$(A\otimes B)|v\rangle$ is $AVB^T$. For
$X\in M_d(\mathbb C)$, define
$|\phi_X\rangle_j:=
(X\otimes I_{B_j})|\Phi_d\rangle_j$, $j=1,2$.
The coefficient matrix of
$|\Phi_d\rangle_{\rm out}\otimes|\chi\rangle$ across
$A'C_A|B'C_B$ is $Q_\chi/\sqrt d$. The two support constraints in
Eq.~(\ref{eq:sm-cat-two-support}) therefore imply the existence of
linear functionals
$\alpha,\beta:M_d(\mathbb C)\to\mathbb C$ such that
\begin{align}
\sum_i
M_i(X\otimes D_\chi)N_i^T
&=
\alpha(X)Q_\chi,
\label{eq:sm-cat-first}
\\
\sum_{r,s}
X_{rs}M_rQ_\chi N_s^T
&=
\beta(X)Q_\chi
\label{eq:sm-cat-second}
\end{align}
for every $X\in M_d(\mathbb C)$, where
$X_{rs}:=\langle r|X|s\rangle$ and $T$ denotes transposition in the
fixed local bases. For $X=I_d$, the left-hand sides of
Eqs.~(\ref{eq:sm-cat-first}) and
(\ref{eq:sm-cat-second}) coincide and are the coefficient matrix of
the nonzero vector in Eq.~(\ref{eq:sm-cat-nonzero}). Hence
$\alpha(I_d)=\beta(I_d)\neq0$.

Second, we show that the matrix constraints obtained above are
incompatible with the nonzero condition.

Let
$E_{rs}:=|r\rangle\langle s|$ denote the standard matrix units and
set $c_{rs}:=\beta(E_{rs})$. Taking $X=E_{rs}$ in
Eq.~(\ref{eq:sm-cat-second}) gives
\begin{eqnarray}
M_rQ_\chi N_s^T
=
c_{rs}Q_\chi.
\label{eq:sm-cat-rank}
\end{eqnarray}
Since $\sum_i c_{ii}=\beta(I_d)\neq0$, there exists an index $t$ such
that $c_{tt}\neq0$. Equation~(\ref{eq:sm-cat-rank}) then gives
$M_tQ_\chi N_t^T=c_{tt}Q_\chi$. Because $Q_\chi$ is invertible, both
$M_t$ and $N_t$ are invertible. Setting $s=t$ and $r=t$,
respectively, in Eq.~(\ref{eq:sm-cat-rank}) yields
\begin{eqnarray}
M_i=a_iM_t,
\qquad
N_i=b_iN_t
\label{eq:sm-cat-proportional}
\end{eqnarray}
for some scalars $a_i,b_i\in\mathbb C$.
Substituting Eq.~(\ref{eq:sm-cat-proportional}) into
Eq.~(\ref{eq:sm-cat-first}) gives
\begin{eqnarray}
\gamma M_t(X\otimes D_\chi)N_t^T
=
\alpha(X)Q_\chi,
\qquad
\gamma:=\sum_i a_ib_i.
\label{eq:sm-cat-rigidity}
\end{eqnarray}
Taking $X=I_d$ gives
$\gamma c_{tt}=\alpha(I_d)\neq0$. Moreover,
$N_t^T=c_{tt}Q_\chi^{-1}M_t^{-1}Q_\chi$.
Substituting this expression into
Eq.~(\ref{eq:sm-cat-rigidity}) and using
$(X\otimes D_\chi)Q_\chi^{-1}=X\otimes I_{R_\chi}$ gives
$\gamma c_{tt}
M_t(X\otimes I_{R_\chi})M_t^{-1}Q_\chi
=\alpha(X)Q_\chi$.
Right-multiplying by $Q_\chi^{-1}$, then left-multiplying by
$M_t^{-1}$ and right-multiplying by $M_t$, yields
\begin{eqnarray}
X\otimes I_{R_\chi}
=
\frac{\alpha(X)}{\gamma c_{tt}}
I_{dR_\chi}
\label{eq:sm-cat-contradiction}
\end{eqnarray}
for every $X\in M_d(\mathbb C)$. Choosing
$X=|0\rangle\langle1|$ contradicts
Eq.~(\ref{eq:sm-cat-contradiction}) for $d\geq2$.
\hfill$\blacksquare$

Using the lemma and the constructions in the proofs of
Theorems~1 and 2, we obtain the following theorem.

\textbf{Theorem 3.} Universal EPPs remain impossible under LOCC on
the full bipartite and multipartite entangled-state sets
considered in Theorems~1 and 2, even with the assistance of pure catalyst.

\textit{Proof.} (a) We first consider the bipartite case with $d_1=d_2$. Assume, for contradiction, that there exist a
finite-dimensional pure catalyst
$|\chi\rangle_{C_AC_B}$ and an $n$-to-$1$ LOCC EPP assisted by this
catalyst that is universal for $\mathcal S_d$.
For each input pair $A_jB_j$, define
$\tau_j=p\Phi_d^{(j)}+(1-p)I_{A_jB_j}/d^2$, where
$1/(d+1)<p<1$. As in the proof of Theorem~1, every $\tau_j$ is full
rank and entangled. Consider
$\Omega_0:=\bigotimes_{j=1}^n\Phi_d^{(j)}$ and
$\Omega_k:=\Phi_d^{(k)}\otimes\bigotimes_{j\neq k}\tau_j$ for
$k=1,\ldots,n$. Since each $\Omega_k$ contains a factor $\Phi_d$, universality and exact catalyst return require
\begin{eqnarray}
\Gamma_0(\Omega_k\otimes\chi)
=
q_k\Phi_d^{\rm out}\otimes\chi,
\quad
q_k>0,
\quad
k=0,\ldots,n.
\label{eq:sm-cat-pure-output}
\end{eqnarray}
Write the successful branch in a product Kraus representation as
$\Gamma_0(\rho)=\sum_\mu K_\mu\rho K_\mu^\dagger$, where
$K_\mu=M_\mu\otimes N_\mu$, and define
$P_{\rm cat}^{\perp}:=
I_{A'B'C_AC_B}-\Phi_d^{\rm out}\otimes\chi$.
For $k=1,\ldots,n$,
Eq.~(\ref{eq:sm-cat-pure-output}) gives
\begin{eqnarray}
P_{\rm cat}^{\perp}
\Gamma_0(\Omega_k\otimes\chi)
P_{\rm cat}^{\perp}
=
\sum_\mu
P_{\rm cat}^{\perp}K_\mu
(\Omega_k\otimes\chi)
K_\mu^\dagger P_{\rm cat}^{\perp}=0.
\label{eq:sm-cat-positive-sum}
\end{eqnarray}
Since every summand in Eq.~(\ref{eq:sm-cat-positive-sum}) is positive
semidefinite and their sum is zero, each summand must vanish. As $\chi$ is a pure state, it follows that
$P_{\rm cat}^{\perp}K_\mu
(\Omega_k^{1/2}\otimes\chi)=0$
for every $\mu$ and $k=1,\ldots,n$.
Because every $\tau_j$ is full rank, we obtain
$\operatorname{supp}\Omega_k=
\operatorname{span}\{|\Phi_d\rangle_k\}
\otimes\mathcal H_{\bar k}^{\rm in}$, where
$\mathcal H_{\bar k}^{\rm in}:=
\bigotimes_{j\neq k}
(\mathcal H_{A_j}\otimes\mathcal H_{B_j})$.
It follows that
\begin{eqnarray}
K_\mu
\left(
|\Phi_d\rangle_k\otimes|\psi\rangle_{\bar k}
\otimes|\chi\rangle
\right)
\in
\mathcal L_{\rm cat}
\label{eq:sm-cat-support}
\end{eqnarray}
for every $\mu$, every $k=1,\ldots,n$, and every
$|\psi\rangle_{\bar k}\in\mathcal H_{\bar k}^{\rm in}$.
The condition $q_0>0$ ensures that at least one product Kraus
operator, denoted by $K=M\otimes N$, acts nontrivially on
$\bigotimes_{j=1}^n|\Phi_d\rangle_j\otimes|\chi\rangle$.
We next reduce $K$ to a product operator acting only on the first two
input pairs and the catalyst, as required to apply the lemma.
For $n=2$, we simply set $\tilde K=K$, and no reduction is required.
For $n>2$, let $\mathcal R=\{3,\ldots,n\}$ denote the set of indices
of the remaining input pairs. For a multi-index
$\mathbf z=(z_3,\ldots,z_n)\in\{0,\ldots,d-1\}^{n-2}$, define
$|\mathbf z\rangle_{A_{\mathcal R}}
:=\bigotimes_{j=3}^n|z_j\rangle_{A_j}$ and
$|\mathbf z\rangle_{B_{\mathcal R}}
:=\bigotimes_{j=3}^n|z_j\rangle_{B_j}$. The product of the target
states on the remaining input pairs can be expanded as
$\bigotimes_{j=3}^n|\Phi_d\rangle_j
=d^{-(n-2)/2}\sum_{\mathbf z}
|\mathbf z\rangle_{A_{\mathcal R}}
|\mathbf z\rangle_{B_{\mathcal R}}$.
Expanding the remaining target states in the nonzero vector
$K(\bigotimes_{j=1}^n|\Phi_d\rangle_j\otimes|\chi\rangle)$ gives a
nonzero sum. Hence there exists at least one multi-index
$\mathbf z_0$ for which the corresponding term is nonzero. Define
$\tilde M$ and $\tilde N$ by fixing the inputs of $M$ and $N$ on the
remaining input pairs to
$|\mathbf z_0\rangle_{A_{\mathcal R}}$ and
$|\mathbf z_0\rangle_{B_{\mathcal R}}$, respectively, and set
$\tilde K:=\tilde M\otimes\tilde N$. The resulting product operator
acts on the first two input pairs and the catalyst.
In either case, $\tilde K$ satisfies the nonzero condition in
Eq.~(\ref{eq:sm-cat-nonzero}), while
Eq.~(\ref{eq:sm-cat-support}) implies the two support constraints in
Eq.~(\ref{eq:sm-cat-two-support}). This contradicts the lemma. Hence
no universal catalytic LOCC EPP exists for $\mathcal S_d$.

We next consider the $d_1\neq d_2$ case.

Set $d=\min\{d_1,d_2\}$ and, for each
$j=1,\ldots,n$, let
$\tau_j=p\Phi_d^{(j)}
+(1-p)I_{A_jB_j}/(d_1d_2)$, where $1/(d+1)<p<1$.
Let $\Omega_0:=\bigotimes_{j=1}^n\Phi_d^{(j)}$ and
$\Omega_k:=\Phi_d^{(k)}\otimes\bigotimes_{j\neq k}\tau_j$ for
$k=1,\ldots,n$. Since $1-p>0$ and $\tau_j^{T_{B_j}}$ has an eigenvalue
$(1-p)/(d_1d_2)-p/d<0$, each $\tau_j$ is full rank and entangled. Consequently,
$\Omega_0,\ldots,\Omega_n$ are allowed total input states.
The positivity, Kraus-selection, and reduction arguments used above
for the case $d_1=d_2$ yield a product operator
$\tilde K=\tilde M\otimes\tilde N$ acting on the first two
input pairs and the catalyst such that
$\tilde K(
|\Phi_d\rangle_1\otimes|\Phi_d\rangle_2\otimes|\chi\rangle)
=c|\Phi_d\rangle_{\rm out}\otimes|\chi\rangle$
for some $c\neq0$. Moreover,
$\tilde K(
|\Phi_d\rangle_1\otimes|\psi\rangle_2\otimes|\chi\rangle)
\in\mathcal L_{\rm cat}$ and
$\tilde K(
|\psi\rangle_1\otimes|\Phi_d\rangle_2\otimes|\chi\rangle)
\in\mathcal L_{\rm cat}$
for every
$|\psi\rangle_j\in
\mathcal H_{A_j}\otimes\mathcal H_{B_j}$, $j=1,2$.
For $j=1,2$, let
$\mathcal K_{A_j}:=
\operatorname{span}\{|r\rangle_{A_j}\}_{r=0}^{d-1}$ and
$\mathcal K_{B_j}:=
\operatorname{span}\{|r\rangle_{B_j}\}_{r=0}^{d-1}$,
and let $P_{A_j}$ and $P_{B_j}$ denote the projectors onto these
subspaces, respectively. For the output subsystems $A'$ and $B'$, let
$\mathcal K_{A'}:=
\operatorname{span}\{|r\rangle_{A'}\}_{r=0}^{d-1}$ and
$\mathcal K_{B'}:=
\operatorname{span}\{|r\rangle_{B'}\}_{r=0}^{d-1}$,
and let $P_{A'}$ and $P_{B'}$ denote the projectors onto these
subspaces, respectively. Define
$\widehat M:=
(P_{A'}\otimes I_{C_A})\tilde M
(P_{A_1}\otimes P_{A_2}\otimes I_{C_A})$
and
$\widehat N:=
(P_{B'}\otimes I_{C_B})\tilde N
(P_{B_1}\otimes P_{B_2}\otimes I_{C_B})$.
By construction, the input projections act as the identity on
$\mathcal K_{A_1}\otimes\mathcal K_{A_2}\otimes\mathcal H_{C_A}$
and
$\mathcal K_{B_1}\otimes\mathcal K_{B_2}\otimes\mathcal H_{C_B}$,
respectively. Since
$|\Phi_d\rangle_{\rm out}\in
\mathcal K_{A'}\otimes\mathcal K_{B'}$, the output projection
$(P_{A'}\otimes I_{C_A})\otimes
(P_{B'}\otimes I_{C_B})$
leaves every vector in $\mathcal L_{\rm cat}$ unchanged. Consequently, $\widehat K:=\widehat M\otimes\widehat N$ is nonzero on
$|\Phi_d\rangle_1\otimes|\Phi_d\rangle_2\otimes|\chi\rangle$ and
satisfies two support conditions of the form in
Eq.~(\ref{eq:sm-cat-support}), with the arbitrary input vector
restricted to
$\mathcal K_{A_j}\otimes\mathcal K_{B_j}$ for $j=1,2$.
Each of the subspaces
$\mathcal K_{A_1}$, $\mathcal K_{A_2}$,
$\mathcal K_{B_1}$, $\mathcal K_{B_2}$,
$\mathcal K_{A'}$, and $\mathcal K_{B'}$
is $d$-dimensional and may therefore be identified with
$\mathbb C^d$. Under these identifications, $\widehat K$ satisfies
Eqs.~(\ref{eq:sm-cat-nonzero}) and
(\ref{eq:sm-cat-two-support}). This contradicts the lemma and proves
the bipartite statement for all finite $d_1,d_2\geq2$.

(b) We then consider the multipartite case.

Suppose, for contradiction, that a catalytic LOCC EPP is
universal for $\mathcal G_{\boldsymbol d}$. Let us consider the states
$\tau_j,\Theta_0$, and $\Theta_k$ introduced in the proof of
Theorem~2. Every $\tau_j$ is full rank and genuinely multipartite
entangled, so all $\Theta_k$, with k=0,\ldots,n are allowed input states. Since each \(\Theta_k\) contains a factor
\(\Psi_{\boldsymbol d}\), universality and exact catalyst
return require
\begin{eqnarray}
\Gamma_0(\Theta_k\otimes\chi)
=
q_k\Psi_{\boldsymbol d}^{\rm out}\otimes\chi,
\quad q_k>0,
\quad k=0,\ldots,n.
\end{eqnarray}
Repeating the positivity, Kraus-selection, and reduction arguments
above gives a multipartite product operator
$\tilde K=\bigotimes_{X\in\mathcal P}\tilde M_X$
acting on the first two input factors and the catalyst. This operator acts nontrivially on $|\Psi_{\boldsymbol d}\rangle_1\otimes |\Psi_{\boldsymbol d}\rangle_2\otimes|\chi\rangle$ and maps both $|\Psi_{\boldsymbol d}\rangle_1\otimes|\phi\rangle_2 \otimes|\chi\rangle$ and $|\phi\rangle_1\otimes|\Psi_{\boldsymbol d}\rangle_2 \otimes|\chi\rangle$ into $\mathcal L_{\rm cat}^{\rm multi}:= \operatorname{span}\{ |\Psi_{\boldsymbol d}\rangle_{\rm out}\otimes|\chi\rangle\}$ for every $|\phi\rangle_j$, $j=1,2$.
Choose a party $A\in\mathcal P$ and write
$\bar A:=\mathcal P\setminus\{A\}$. Let
$\mathcal H_{C_{\bar A}}:=
\bigotimes_{X\in\bar A}\mathcal H_{C_X}$, and let
$I_{C_{\bar A}}$ denote the identity operator on
$\mathcal H_{C_{\bar A}}$. Use the
$d$-dimensional input and output subspaces
$\mathcal U_1,\mathcal U_2,\mathcal U_{\rm out}$ and
$\mathcal V_1,\mathcal V_2,\mathcal V_{\rm out}$ introduced in the
proof of Theorem~2. Let $P_{\mathcal U_\ell}$ and $P_{\mathcal V_\ell}$ denote the
projectors onto the corresponding input subspaces, and let
$P_{\mathcal U_{\rm out}}$ and $P_{\mathcal V_{\rm out}}$ denote the
output projectors. Across the bipartition $A|\bar A$, write
$\tilde K=
\tilde M_A\otimes\tilde N_{\bar A}$, where
$\tilde N_{\bar A}:=
\bigotimes_{X\in\bar A}\tilde M_X$.
 Define
$\widehat M:=
(P_{\mathcal U_{\rm out}}\otimes I_{C_A})
\widetilde M_A
(P_{\mathcal U_1}\otimes P_{\mathcal U_2}\otimes I_{C_A})$
and
$\widehat N:=
(P_{\mathcal V_{\rm out}}\otimes I_{C_{\bar A}})
\widetilde N_{\bar A}
(P_{\mathcal V_1}\otimes P_{\mathcal V_2}
\otimes I_{C_{\bar A}})$.
The input projections $P_{\mathcal U_1}\otimes P_{\mathcal U_2}\otimes I_{C_A}$ and $P_{\mathcal V_1}\otimes P_{\mathcal V_2}
\otimes I_{C_{\bar A}}$ restrict the two arbitrary input states to the
chosen subspaces. Since every resulting output is proportional to
$|\Psi_{\boldsymbol d}\rangle_{\rm out}\otimes|\chi\rangle$, and
$|\Psi_{\boldsymbol d}\rangle_{\rm out}$ belongs to
$\mathcal U_{\rm out}\otimes\mathcal V_{\rm out}$, the output
projections $P_{\mathcal U_{\rm out}}\otimes I_{C_A}$ and $P_{\mathcal V_{\rm out}}\otimes I_{C_{\bar A}}$ leave these states unchanged.
Under the identifications used in the proof of Theorem~2, each
target state $|\Psi_{\boldsymbol d}\rangle_{\rm out}$ becomes a maximally entangled state of
Schmidt rank $d$ across $A|\bar A$. Moreover,
$\widehat K:=\widehat M\otimes\widehat N$ is a product operator across
$AC_A\,|\,\bar A C_{\bar A}$ and remains nonzero on $|\Psi_{\boldsymbol d}\rangle_1\otimes |\Psi_{\boldsymbol d}\rangle_2\otimes|\chi\rangle$. It therefore satisfies
Eqs.~(\ref{eq:sm-cat-nonzero}) and
(\ref{eq:sm-cat-two-support}), with the two sides of the bipartition
treated as effective parties. This contradicts the lemma.

Hence no universal catalytic LOCC EPP exists for either
$\mathcal S_{d_1,d_2}$ or $\mathcal G_{\boldsymbol d}$.
\hfill$\blacksquare$

\section*{SIII. Universal coherence purification under strictly incoherent operations}

We first recall the basic notions of coherence purification under
strictly incoherent operations~\cite{Baumgratz2014,Winter2016,Yadin2016}. Fix an orthonormal basis
$\{|r\rangle\}_{r=0}^{d-1}$ of $\mathbb C^d$, called the incoherent
basis. A state $\delta$ on $\mathbb C^d$ is incoherent if and only if
it can be written as
$\delta=\sum_{r=0}^{d-1}\delta_r|r\rangle\langle r|$, where
$\delta_r\geq0$ and $\sum_{r=0}^{d-1}\delta_r=1$. A state that is not
diagonal in the incoherent basis is called coherent. We denote the
sets of incoherent and coherent states on $\mathbb C^d$ by
$\mathcal I_d$ and $\mathcal C_d$, respectively. Relative to this basis, we define the maximally coherent state by
$|\varphi_d\rangle=d^{-1/2}\sum_{r=0}^{d-1}|r\rangle$ and write
$\varphi_d:=|\varphi_d\rangle\langle\varphi_d|$.

A strictly incoherent operation (SIO) from
$\mathcal H_{\rm in}$ to $\mathcal H_{\rm out}$ is a completely
positive and trace-preserving map $\mathcal E$ that admits a Kraus
representation
$\mathcal E(\rho)=\sum_\mu K_\mu\rho K_\mu^\dagger$
in which every $K_\mu$ is strictly incoherent. Here,
$\mathcal H_{\rm in}$ and $\mathcal H_{\rm out}$ are equipped with
incoherent bases
$\{|r\rangle_{\rm in}\}_{r=0}^{d_{\rm in}-1}$ and
$\{|s\rangle_{\rm out}\}_{s=0}^{d_{\rm out}-1}$, respectively. A
linear operator
$K:\mathcal H_{\rm in}\to\mathcal H_{\rm out}$ is strictly incoherent
if
$K|r\rangle_{\rm in}
=\alpha_r|f(r)\rangle_{\rm out}$ and
$K^\dagger|s\rangle_{\rm out}
=\beta_s|g(s)\rangle_{\rm in}$
for every $r$ and $s$, where
$\alpha_r,\beta_s\in\mathbb C$ may be zero, and $f(r)$ and $g(s)$ are
output and input incoherent basis labels, respectively. Equivalently,
the matrix of $K$ with respect to these bases has at most one nonzero
entry in each row and each column
~\cite{Winter2016,Yadin2016}.

An $n$-to-$1$ SIO coherence purification protocol is a two-outcome
quantum instrument $\Lambda=\{\Lambda_0,\Lambda_1\}$ consisting of a
success branch $\Lambda_0$ and a failure branch $\Lambda_1$, each of
which is a completely positive and trace-nonincreasing map from
$\mathcal B((\mathbb C^d)^{\otimes n})$ to
$\mathcal B(\mathbb C^d)$ and admits a Kraus representation consisting
of strictly incoherent operators, with $\Lambda_0+\Lambda_1$ being
trace preserving. Here,
the input incoherent basis is the tensor product of the incoherent
bases of the $n$ input factors, while the output incoherent basis is
$\{|r\rangle\}_{r=0}^{d-1}$.

An $n$-to-$1$ SIO coherence purification protocol
$\Lambda=\{\Lambda_0,\Lambda_1\}$ is universal for $\mathcal C_d$ if,
for every total input state
$\rho^{\mathrm{in}}=\rho_1\otimes\cdots\otimes\rho_n$ with
$\rho_j\in\mathcal C_d$ for all $j=1,\ldots,n$, the success probability
$p_{\rm s}:=\operatorname{Tr}[\Lambda_0(\rho^{\mathrm{in}})]$ is
positive and the normalized output conditioned on success,
$\rho':=\Lambda_0(\rho^{\mathrm{in}})/p_{\rm s}$, satisfies
\begin{eqnarray}
F_{\rm c}(\rho')
\geq
\max_{1\leq j\leq n}F_{\rm c}(\rho_j).
\label{eq:sm-coherence-universality}
\end{eqnarray}
Here,
$F_{\rm c}(\rho):=\langle\varphi_d|\rho|\varphi_d\rangle$
denotes the fidelity of a state $\rho$ on $\mathbb C^d$ with the
maximally coherent target state $|\varphi_d\rangle$.

For $\epsilon>0$, define the restricted set
\begin{eqnarray}
\mathcal C_d^{(\epsilon)}
:=
\{\rho\in\mathcal C_d:F_{\rm c}(\rho)>1-\epsilon\}.
\label{eq:sm-high-coherence-set}
\end{eqnarray}

\textbf{Theorem 4.} For every $d\geq2$ and any $n\geq2$,
no $n$-to-$1$ SIO coherence-purification protocol is universal for
$\mathcal C_d$. Moreover, for every $\epsilon>0$, no such protocol is
universal for $\mathcal C_d^{(\epsilon)}$.

\textit{Proof.} (a) We prove the statement for $\mathcal C_d$.
Suppose, for contradiction, that a universal $n$-to-$1$ SIO
coherence-purification protocol exists.

First, we construct total input states for which universality requires
the normalized output conditioned on success to be the maximally
coherent state.

Consider the $n+1$ total input states
$\Sigma_0:=\bigotimes_{j=1}^n\varphi_d^{(j)}$ and
$\Sigma_k:=\varphi_d^{(k)}
\otimes\bigotimes_{j\neq k}\sigma_j$ for $k=1,\ldots,n$.
Here, for each $j=1,\ldots,n$, let
$|\varphi_d\rangle_j
=d^{-1/2}\sum_{r=0}^{d-1}|r\rangle_j$
denote the maximally coherent state on the $j$th input Hilbert space
and set
$\varphi_d^{(j)}:=|\varphi_d\rangle_j\langle\varphi_d|$.
For each $j=1,\ldots,n$, let
\begin{eqnarray}
\sigma_j
&:=&
p\varphi_d^{(j)}
+
(1-p)\frac{I_j}{d},
\qquad
0<p<1,
\label{eq:sm-coherence-sigma}
\end{eqnarray}
where $I_j$ is the identity on the $j$th input Hilbert space.
Since $0<p<1$, every $\sigma_j$ is full rank and coherent. Thus,
$\Sigma_0,\ldots,\Sigma_n$ are allowed total input states. Let $|\varphi_d\rangle_{\rm out}$ denote the maximally coherent target
state on the output Hilbert space and set
$\varphi_d^{\rm out}:=
|\varphi_d\rangle_{\rm out}\langle\varphi_d|$. Moreover,
each of $\Sigma_k$ contains at least one maximally coherent target state, so
the maximum input fidelity in
Eq.~(\ref{eq:sm-coherence-universality}) is one. Universality therefore
requires
\begin{eqnarray}
\Lambda_0(\Sigma_k)
&=&
p_k\varphi_d^{\rm out},
\qquad
k=0,\ldots,n,
\label{eq:sm-coherence-pure-output}
\end{eqnarray}
where
$p_k:=\operatorname{Tr}[\Lambda_0(\Sigma_k)]>0$.

Second, we derive a support constraint and select a Kraus operator that
acts nontrivially on $\Sigma_0$.

By definition of an SIO coherence-purification protocol, the
successful branch $\Lambda_0$ admits a Kraus representation
$\Lambda_0(\rho)=\sum_\mu K_\mu\rho K_\mu^\dagger$ in which every
$K_\mu$ is strictly incoherent. Let
$\mathcal L_{\rm c}:=
\operatorname{span}\{|\varphi_d\rangle_{\rm out}\}$, and let
$P_\perp:=I_{\rm out}-\varphi_d^{\rm out}$ denote the orthogonal
projector onto $\mathcal L_{\rm c}^{\perp}$, where $I_{\rm out}$ is
the identity on the output Hilbert space. For $k=1,\ldots,n$,
Eq.~(\ref{eq:sm-coherence-pure-output}) gives
\begin{eqnarray}
P_\perp\Lambda_0(\Sigma_k)P_\perp
=
\sum_\mu
P_\perp K_\mu\Sigma_kK_\mu^\dagger P_\perp=0 .
\label{eq:sm-coherence-positive-sum}
\end{eqnarray}
All summands in Eq.~(\ref{eq:sm-coherence-positive-sum}) are positive
semidefinite, and their sum is zero; hence each summand must vanish.
It follows that $P_\perp K_\mu\Sigma_k^{1/2}=0$ for every $\mu$ and
$k=1,\ldots,n$. Since every $\sigma_j$ is full rank, the range of
$\Sigma_k^{1/2}$ is
$\operatorname{supp}\Sigma_k=
\operatorname{span}\{|\varphi_d\rangle_k\}\otimes
\mathcal H_{\bar k}^{\rm c}$, where
$\mathcal H_{\bar k}^{\rm c}:=
\bigotimes_{j\neq k}\mathbb C^d$ is the Hilbert space of all input
factors except the $k$th; consequently,
$P_\perp K_\mu\Sigma_k^{1/2}=0$ implies
\begin{eqnarray}
K_\mu
\left(
|\varphi_d\rangle_k
\otimes|\phi\rangle_{\bar k}
\right)
\in
\mathcal L_{\rm c}
\label{eq:sm-coherence-support}
\end{eqnarray}
for every $\mu$, every $k=1,\ldots,n$, and every
$|\phi\rangle_{\bar k}\in\mathcal H_{\bar k}^{\rm c}$.
For the pure input state $\Sigma_0$,
Eq.~(\ref{eq:sm-coherence-pure-output}) becomes
$\sum_\mu|v_\mu\rangle\langle v_\mu|
=p_0\varphi_d^{\rm out}$, where
$|v_\mu\rangle:=K_\mu|\varphi_d\rangle^{\otimes n}$.
The condition $p_0>0$ ensures that at least one $|v_\mu\rangle$ is
nonzero. We may therefore choose an SIO Kraus operator, denoted by
$K$, such that
\begin{eqnarray}
K|\varphi_d\rangle^{\otimes n}
\neq0 .
\label{eq:sm-coherence-nonzero}
\end{eqnarray}
The selected $K$ also satisfies
Eq.~(\ref{eq:sm-coherence-support}) for every $k=1,\ldots,n$.

Third, we show that the nonzero condition and the support constraints
are incompatible with the row-and-column structure of an SIO Kraus
operator.

Label the incoherent basis of the total input Hilbert space by strings
$\mathbf x=(x_1,\ldots,x_n)\in\{0,\ldots,d-1\}^n$, with
$|\mathbf x\rangle:=
|x_1\rangle_1\otimes\cdots\otimes|x_n\rangle_n$.
The selected Kraus operator can be written as
$K=\sum_{\mathbf x\in S_K}
c_{\mathbf x}|f(\mathbf x)\rangle_{\rm out}\langle\mathbf x|$,
where $S_K$ is the set of labels of the nonzero columns of $K$,
$c_{\mathbf x}\neq0$ for every $\mathbf x\in S_K$, and
$f:S_K\to\{0,\ldots,d-1\}$ assigns the output row occupied by each
nonzero column. Because every output row contains at most one nonzero
entry, $f$ is injective. Consequently, the cardinality of $S_K$, denoted by $|S_K|$, satisfies
\begin{eqnarray}
|S_K|\leq d .
\label{eq:sm-S-bound}
\end{eqnarray}
Equation~(\ref{eq:sm-coherence-nonzero}) implies that $S_K$ is
nonempty. Fix a string
$\mathbf x=(x_1,\ldots,x_n)\in S_K$, which we call the selected
string. Keeping its last $n-1$ components fixed, consider the vector
$|\varphi_d\rangle_1\otimes|x_2,\ldots,x_n\rangle$, where
$|x_2,\ldots,x_n\rangle:=
\bigotimes_{j=2}^n|x_j\rangle_j$.
This vector has support only on the $d$ basis strings
$(r,x_2,\ldots,x_n)$, with $r=0,\ldots,d-1$, and contains the
selected string $\mathbf x$. Its image under $K$ is nonzero because
the column labeled by $\mathbf x$ contributes a nonzero component,
whereas distinct nonzero columns occupy distinct output rows and
therefore cannot cancel one another.
By Eq.~(\ref{eq:sm-coherence-support}), this nonzero output must be
proportional to $|\varphi_d\rangle_{\rm out}$ and hence has the form
$c|\varphi_d\rangle_{\rm out}
=(c/\sqrt d)\sum_{r=0}^{d-1}|r\rangle_{\rm out}$ for some $c\neq0$.  Since each output
row of $K$ contains at most one nonzero entry, all $d$ strings
$(r,x_2,\ldots,x_n)$ must belong to $S_K$. Together with
Eq.~(\ref{eq:sm-S-bound}), this gives
\begin{eqnarray}
S_K
=
\{(r,x_2,\ldots,x_n):0\leq r\leq d-1\}.
\label{eq:sm-S-first}
\end{eqnarray}
We next apply Eq.~(\ref{eq:sm-coherence-support}) with $k=2$ to the vector
$|x_1\rangle_1\otimes|\varphi_d\rangle_2
\otimes|x_3,\ldots,x_n\rangle$, where, for $n=2$, the last tensor
factor is omitted. This vector contains the same selected basis string
$\mathbf x\in S_K$, so its image under $K$ is again nonzero.
Repeating the preceding argument gives
\begin{eqnarray}
S_K
=
\{(x_1,r,x_3,\ldots,x_n):0\leq r\leq d-1\}.
\label{eq:sm-S-second}
\end{eqnarray}
For $d\geq2$, Eqs.~(\ref{eq:sm-S-first}) and
(\ref{eq:sm-S-second}) are incompatible: the set in
Eq.~(\ref{eq:sm-S-first}) contains a string whose first coordinate
$r$ differs from $x_1$, whereas every string in
Eq.~(\ref{eq:sm-S-second}) has first coordinate $x_1$. This
contradiction proves that no $n$-to-$1$ SIO
coherence-purification protocol is universal for $\mathcal C_d$.

(b) It remains to prove the statement for
$\mathcal C_d^{(\epsilon)}$. The states in
Eq.~(\ref{eq:sm-coherence-sigma}) satisfy
$F_{\rm c}(\sigma_j)=p+(1-p)/d$, which approaches one as $p\to1$.
For every fixed $\epsilon>0$, we may therefore choose $p<1$
sufficiently close to one so that every $\sigma_j$ belongs to
$\mathcal C_d^{(\epsilon)}$. The  state
$\varphi_d$ also belong to $\mathcal C_d^{(\epsilon)}$ and have
unit fidelity. Hence all total input states used in the preceding
contradiction can be constructed from states in
$\mathcal C_d^{(\epsilon)}$. The same contradiction proves that no
$n$-to-$1$ SIO coherence-purification protocol is universal for
$\mathcal C_d^{(\epsilon)}$.
\hfill$\blacksquare$


\begin{thebibliography}{99}
\bibitem{note2}
Here, universality does not mean the complete absence of prior
information; rather, it means that the protocol depends only on
general information about the inputs, such as whether they are pure
or mixed, their local dimensions, the number of parties, and whether
they are independently and identically distributed (i.i.d.).
\end{thebibliography}
\end{document}